# Cardiac Arrhythmia Detection from ECG Combining Convolutional and Long Short-Term Memory Networks

Philip Warrick[1], Masun Nabhan Homsi[2]

[1]PeriGen. Inc., Montreal, Canada
[2]Simon Bolivar University, Caracas, Venezuela

**Abstract**

*Objectives: Atrial fibrillation (AF) is a common heart rhythm disorder associated with deadly and debilitating consequences including heart failure, stroke, poor mental health, reduced quality of life and death. Having an automatic system that diagnoses various types of cardiac arrhythmias would assist cardiologists to initiate appropriate preventive measures and to improve the analysis of cardiac disease. To this end, this paper introduces a new approach to detect and classify automatically cardiac arrhythmias in electrocardiograms (ECG) recordings.*

*Methods: The proposed approach used a combination of Convolution Neural Networks (CNNs) and a sequence of Long Short-Term Memory (LSTM) units, with pooling, dropout and normalization techniques to improve their accuracy. The network predicted a classification at every 18th input sample and we selected the final prediction for classification. Results were cross-validated on the Physionet Challenge 2017 training dataset, which contains 8,528 single lead ECG recordings lasting from 9s to just over 60s.*

*Results: Using the proposed structure and no explicit feature selection, 10-fold stratified cross-validation gave an overall F-measure of 0.83.10±0.015 on the held-out test data (mean ± standard deviation over all folds) and 0.80 on the hidden dataset of the Challenge entry server.*

## 1. Introduction

Electrocardiogram (ECG) represents electrical activity of human heart. It is composite from 5 waves: P, Q, R, S and T. P wave is the Atria depolarization (Atrial Contraction), T wave is the depolarization of ventricles (Ventricular relaxation). QRS complex represents ventricles depolarization (Ventricular contraction). The QRS complex, ST segment, PR interval, RR interval, PR segment, QT interval are the most important sections in an ECG signal for the diagnosis of different cardiac diseases, especially arrhythmia. An arrhythmia is an alteration of the regular rate or rhythm of the heartbeat. The most common type of arrhythmia is atrial fibrillation (AF), which causes an irregular and fast heartbeat [1].

Automatic ECG classification is an emerging tool for the cardiologists in medical diagnosis for effective treatments. Traditional methods to classify ECG signal include pre-processing, feature extraction and classification steps. Various kinds of noise and artefacts are first removed, and then the signal is segmented to calculate features vectors over time, possibly using a feature reduction algorithm to reduce dimensionality. Finally, for classification the features form the input to a machine learning algorithm such as Support Vector Machines (SVMs) [2], Neural Networks (NNs) [3] or ensemble learning [4].

Deep learning (DL) based neural network models have achieved great success in multiple fields such as natural language processing, computer vision, biomedical signal processing and others. DLs can overcome the challenge of the often tedious feature engineering task and helps with parameterizing traditional NN with many layers [5]. Convolution Neural Networks (CNNs) and Recurrent Neural Networks (RNNs) especially Long Short-term Memory Network (LSTM) are two widely used architectures among these models. CNNs learn appropriate filters that reduce the input dimensionality, while LSTMs are useful to model system dynamics [5].

Many approaches have been performed to classify various cardiac arrhythmias that used DL models. Authors in [6] developed a new model that predicted 12 arrhythmias from single-lead ECG. It consisted of 34-layer CNN that maps a sequence of ECG samples to a sequence of rhythm classes. A new neural network architecture was proposed in [7] for anomaly detection in ECG time signals. It consisted by stacking multiple recurrent LSTM layers. Another approach based on DL for classification of ECG signals was proposed in [8]. It consisted of two layers: feature representation layer and softmax regression layer. The feature representation layer was trained from the raw ECG data in an unsupervised way employing stacked denoising autoencoders with sparsity constraint.

In this paper, we propose a novel deep neural network that combines CNN and LSTM to effectively learn sequence data containing longer term patterns of unknown length extracted from ECG signals. The model does not require explicit pre-processing, but can adaptively discover hidden structures of different ECG entities and automatically learn their dependencies. . The output of a one-layer CNN is fed into a stack of three recurrent LSTM layers. The CNN is constructed on top of the signal vectors from a large corpus of ECG data to learn higher-level representations of PQRST regions. In our approach, the task is formulated as a temporal sequence predicting problem that can be solved under a sequence-to-sequence learning framework. The new model classifies ECG signals into normal sinus rhythm (N), atrial fibrillation (A), an alternative rhythm (O), or noisy (~). We used the PhysioNet Challenge 2017 dataset [9] which consists of 8528 ECG signals sampled at 300 Hz and mostly lasting between 30 and 60 s although there were some records as short as 6 s. We will refer to our proposed deep architecture by CL3 (One CNN and three LSTMs).

The rest of the paper is organized as follows. Detailed descriptions of our proposed deep model are presented in sections 2. Section 3 reports experimental settings and evaluation metrics, and discusses experimental results. Finally, conclusions and future directions are outlined in section 4.

## 2. CL3 Model

The architecture of the CL3 model is illustrated in Figure 1, which consists of two main components: one is representation learning, CNN, and the other one is the sequence learning, a stack of three LSTMs. After applying one-layer CNN on the raw input sequence to extract local and discriminative features, three layers of LSTMs are put up on top of the previous CNN to encode the sequential patterns. Then, one dense layer is added to process the output of the third LSTM. Finally, a softmax function is adopted to predict the class.

### 2.1. Model architecture

#### 2.1.1. Representation learning

This component consists of CNN and one max-pooling layer. The convolution layer performs three operations sequentially: 1D-convolution with its filters, batch normalization, with rectified linear unit (ReLU) activation. Pooling layer downsamples the inputs using the max operation. There are two reasons to use a max-pooling layer here. First, it reduces computation for upper layers. Second, it can extract local dependencies and keep the most salient information. The obtained vectors are then fed to the second component which is sequence learning. We used 10 filters to be comparative to other filtering schemes (e.g., the use of up to 10 PCA or wavelet components is typical for biomedical signals). All system parameters including the number of filters and pooling size can be found in section 3. Dropout and Batch Normalization blocks are also explained later in section 2.2. Each ECG recording is considered a univariate time series and it is denoted by X={$x_1$, $x_2$,…,$x_N$}, where N represents the length of the ECG signal. CNN extracts the i-th feature $a_i$ from the i-th ECG sample $x_i$ as follows:

$$a_i = CNN_\theta(x_i) \quad (1)$$

where $CNN_\theta(x_i)$ is a function that transforms an ECG signal into a feature vector $a_i$ using a CNN with θ parameter to represent the number of filters. These features vectors {$a_1$, $a_2$,…, $a_N$} are then forwarded to the sequence learning component.

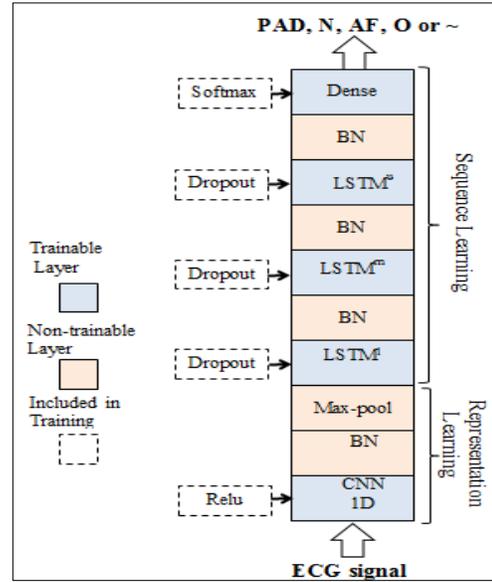

Figure 1. CL3 architecture.

#### 2.1.2. Sequence learning

This component is created by stacking multiple LSTM hidden layers on top of each other, with the output sequence of one layer forming the input sequence for the next. Here three LSTM layers are used. Specifically, input of the upper LSTM layer (*u*) takes $h^m(t)$ from the middle LSTM layer (*m*), and again the middle LSTM layer takes $h^l(t)$ from the lower LSTM layer (*l*). The outputs of the second layer and the third layer are as same operation as $LSTM^l$ and $LSTM^m$.

Formally, suppose there are N features obtained from the CNN {$a_1$, $a_2$,…,$a_N$} organized sequentially and t=1…N denotes the time index of ECG samples, the sequence learning component is defines as follows:

$$h_t^l, c_t^1 = LSTM_{\theta_l}^l(h_{t-1}^l, c_{t-1}^l, a_t) \quad (2)$$
$$h_t^m, c_t^m = LSTM_{\theta_m}^m(h_{t-1}^m, c_{t-1}^m, h_t^l) \quad (3)$$
$$h_t^u, c_t^u = LSTM_{\theta_u}^u(h_{t-1}^u, c_{t-1}^u, h_t^m) \quad (4)$$

where LSTM denotes a function that processes sequences of features at using a stacked of three LSTMs parametrized by $\theta_l$, $\theta_m$ and $\theta_u$ for lower, middle and upper layers; h and c are vectors of hidden and cell states of LSTMs; $h_0^l, c_0^l, h_{N+1}^u$ and $c_{N+1}^u$ are set to zero vectors.

A dense layer forms the final layer and its output is passed to a softmax function whose output is the probability distribution over labels [5].

## 2.2. Learning CL3 for ECG Classification

The entire model is trained by minimizing the cross-entropy error. Given a training sample x and its corresponding label y∈{1,2,…,K} where K is the number of possible labels (ECG classes), and the estimated probabilities $\tilde{y}_j \in [0,1]$ for each label j ∈{1, 2, ..., K}, the error is defined by equation 5, where 1{condition} could be 1 or 0. We used the Root Mean Square Propagation (RMSprop) optimization method that calculates the magnitude of recent gradients to normalize the gradients to optimize model parameters over the following objective function:

$$L(x, y) = \sum_{j=1}^{k} 1\{y = j\} \log(\tilde{y}_j) \qquad (5)$$

Dropout was employed to prevent overfitting during training: it was applied to the input of each LSTM, while Recurrent Dropout was employed to drop neurons directly in recurrent connections of each LSTM Batch Normalization was used to keep values in-bounds and avoid saturation at the various processing steps. It was performed on the outputs of the CNN and all LSTM layers.

Efficient batch-oriented training requires fixed-length input; therefore we defined a maximum length (*max_length*) for the ECG signals in the dataset. Each ECG signal that has a length less than *max_length* was padded with zeros. Zero label is also added to the sequence of labels to be referred as PAD class. On the other hand, those ECG signals that are longer than *max_length* were split into multiple sequences in such that each new sequence length was less than or equal to *max_length*. Because a large majority of signals had 9000 samples (30 s), and a significant number were twice this length (18000), we chose a *max_length* of 9000.

The class vector now contains five different labels [PAD, N, AF, O, ~] which are one-hot encoded. The final target sequence for each input sequence is constructed by repeating the binary vector *rep_length* times, defined by equation 6.

$$rep\_length = max\_length/target\_factor \qquad (6)$$

where *target_factor* is determined experimentally. The final train input shape is (Number of sequences to be trained* *max_length**1), where 1 refers the single input signal dimension, while the final output shape is given by (Number of sequences to be trained * *rep_length* * number of classes).

## 3. Experimental settings and results

Our deep model was implemented in Python using the Keras library with a TensorFlow backend, which provides efficient functionality on CPUs and GPUs. The hyper-parameters were chosen according to experiment results and they are as follows:
*a*) Batch size is 500, *Max_length* of input sequence is 9000, and the weight matrix is initialized to 1. *b*) For CNN layer, both the number of filters and kernel size are equal to 10. *c*) For Max pooling layer, the pool size is 18. *d*) For the three LSTM layers, the number of cells is equal to 100 and both dropout and recurrent dropout are 0.1. *e*) For BN layers, 0.99, 0.001, zeros, ones, zeros, ones are used for momentum, epsilon, beta initializer, gamma initializer, moving mean initializer and moving variance initializer, respectively [5]. *f*) For the dense layer, a normal distribution centered on zero is used for kernel initializer and output shape is (9000, 500, 5). 500 is the result of the *target_factor=18* sample decimation effect of the max pooling layer, while 5 represents number of classes.

The classifier performance was evaluated in terms of F1-measure and it is calculated by F1=2*(P*R)/(P+R). It represents the harmonic mean of Precision (P) and Recall (R). P is the ratio of true positives to all predicted positives and given by P=TP/(TP+FP). For more information, refer to [9].

Table 1. Results of different experiments and entries. σ is the cross-validation F1 standard deviation.

| # | Model | Network variant(s) | 10-folds CV | | | Entry |
|---|-------|--------------------|-------------|---|---|-------|
| | | | Class | F1 (%) | σ (%) | F1 (%) |
| 1 | CL3 | Batch size=500 | N | 90.10 | 0.30 | 92 |
| | | | AF | 76.00 | 6.20 | 85 |
| | | | O | 75.20 | 3.50 | 75 |
| | | | ~ | 47.10 | 10.90 | - |
| **I** | | **Overall** | | **83.10** | **1.50** | **84** |
| 2 | CL3 | Batch size=500 & Class weight | N | 90.20 | 1.00 | 91 |
| | | | AF | 75.90 | 6.50 | 82 |
| | | | O | 75.50 | 3.70 | 74 |
| | | | ~ | 48.50 | 1.34 | - |
| II | | Overall | | 83.30 | 2.20 | 82 |
| 3 | CL3 | Dilated CNN=2 | N | 89.70 | 0.80 | 92 |
| | | | AF | 75.70 | 5.50 | 83 |
| | | | O | 74.10 | 3.40 | 75 |
| | | | ~ | 47.10 | 9.00 | - |
| III | | Overall | | 82.60 | 1.70 | 83 |

Three experiments were conducted to evaluate the effectiveness of the proposed deep architecture for ECG signal classification. Table 1 displays 10-fold cross validation scores of those experiments along with the 3 best entries submitted to the challenge server. It can be noted that Experiment 2 exhibited the best F1 (83.30%), but its standard deviation (σ) (1.80) is higher than

Experiment 1 (1.50). In Experiment 2, imbalanced class issue was dealt by applying weights to misclassification sequences. F1 was better than Experiment 1, but its σ is higher and as result its Entry II performed poorer than the first one with F1 of 82%. Dilated convolution layer was employed in experiment 3 and its results showed that it performed also worse than Experiment 1; yielding lower F1s and higher σ with an entry score that reached up to 83%. We can conclude that Entry I yielded our best overall score of the submitted entries of 83.10±1.50 in training and 82 using the hidden dataset. Our final score after version 3 relabelling was 80.

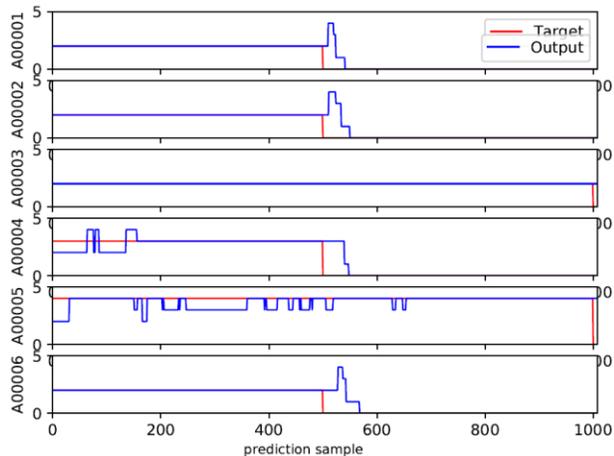

Figure 2. Class prediction vs original class.

Figure 2 depicts 6 sample ECG examples A00001, A00002, A00003, A00004, A00005 and A00006 that were labelled manually by cardiologists as N, N, N, A, A and N, and automatically by our CL3 in Entry I. It can be seen that for A00001, A00002, A00003 and A00006 recordings, CL3 could predict perfectly almost all ECG samples with the same label assigned by the expert, while for A00004 and A0005, there are some ECG samples could not be learnt completely by our new deep architecture. In addition to that, Figure 3 shows F1 results over time per class. It can be noticed from this figure that the model performance begun to degrade for classes ~ and AF, from the middle of training. This is due to the fact that both are the minority classes, while performances of majority classes keep significantly increasing throughout the training (and supporting our decision to use the final prediction for classification).

## 4. Conclusions

A new deep learning model, named CL3, for automatic classification of cardiac arrhythmias based on raw single-lead ECGs is proposed. CL3 uses CNN to extract features which are introduced to a stack of LSTMs to learn automatically hidden patterns from ECG epochs with very little manual parameter tuning required. Our results showed that CL3 is suitable for diagnosis of different cardiac diseases with good accuracy. Future work will focus on refining our model by applying an ensemble deep learning framework to decrease information loss and overfitting problems, and to overcome the class imbalance problem.

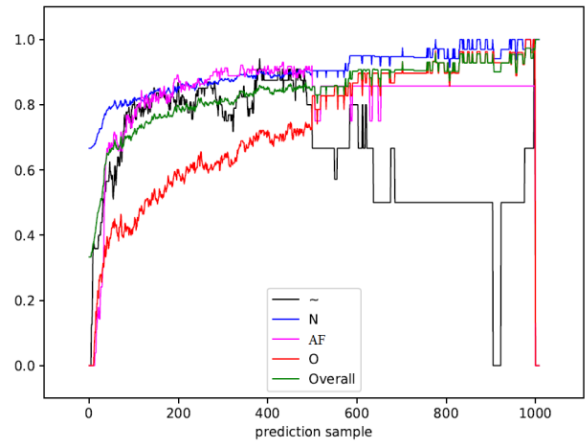

Figure 3. F1-measure curves per class vs training prediction for Entry I.

Address for correspondence.

Philip Warrick, PeriGen. Inc. Montreal, Canada.
philip.warrick@gmail.com
Masun Nabhan Homsi, Universidad Simón Bolívar.
mnabhan@usb.ve